\documentclass[journal]{vgtc}                

\ifpdf
  \pdfoutput=1\relax                   
  \usepackage{graphicx}                
  \DeclareGraphicsExtensions{.pdf,.png,.jpg,.jpeg} 
\else
  \usepackage{graphicx}                
  \DeclareGraphicsExtensions{.eps}     
\fi%

\graphicspath{{figures/}{pictures/}{images/}{./}} 

\PassOptionsToPackage{warn}{textcomp}  
\usepackage{textcomp}                  
\usepackage{mathptmx}                  
\usepackage{times}                     
\usepackage{tabu}                      
\usepackage{booktabs}                  
\usepackage{diagbox}
\usepackage{makecell}

\usepackage{mathptmx}
\usepackage{overpic}
\usepackage{contour}
\usepackage{epstopdf}
\usepackage{caption}
\usepackage{times}
\usepackage[lined,algonl,boxed]{algorithm2e}
\usepackage{amssymb,amsmath}
\usepackage{bbm}
\usepackage{multirow}
\usepackage{subfigure}
\usepackage{paralist}
\usepackage{tikz}
\usepackage{multirow}
\usepackage{color}
\usepackage{setspace}
\usepackage{microtype,hyphenat,balance}
\usepackage{array}
\usepackage{url}
\usepackage{bm}
\usepackage{algorithm2e}
\usepackage{algpseudocode}
\usepackage{amsmath}
\usepackage[implicit=false, bookmarks=false]{hyperref}

\newcommand{\shixia}[1]{\textcolor{black}{#1}}

\newcommand{\jiazhi}[1]{\textcolor{black}{#1}}
\newcommand{\yuanjun}[1]{\textcolor{black}{#1}}
\newcommand{\lingyun}[1]{\textcolor{black}{#1}}
\newcommand{\doc}[1]{\textcolor{black}{#1}}
\newcommand{\revision}[1]{\textcolor{black}{#1}}

\def \etal {{\emph{et al}.\thinspace}}
\def \eg {{\emph{e.g}.\thinspace}}
\def \ie {{\emph{i.e}.\thinspace}}

\onlineid{1062}

\vgtccategory{Research}

\title{Evaluation of Sampling Methods for Scatterplots}
\author{Jun Yuan, Shouxing Xiang, Jiazhi Xia, Lingyun Yu, and Shixia Liu}

\authorfooter{
\item
 J. Yuan, S. Xiang and S. Liu are with Tsinghua University. E-mail: \{yuanj19, xsx18\}@mails.tsinghua.edu.cn, shixia@tsinghua.edu.cn. Shixia Liu is the corresponding author.
\item
 J. Xia is with Central South University. E-mail: xiajiazhi@csu.edu.cn.
\item
 L. Yu is with Xi'an Jiaotong-Liverpool University. E-mail: lingyun.yu@xjtlu.edu.cn.
}

\shortauthortitle{Yuan \MakeLowercase{\textit{et al.}}: Evaluation of Sampling Methods for Scatterplots}

\teaser{
\centering
  \begin{overpic}[width=1\linewidth]{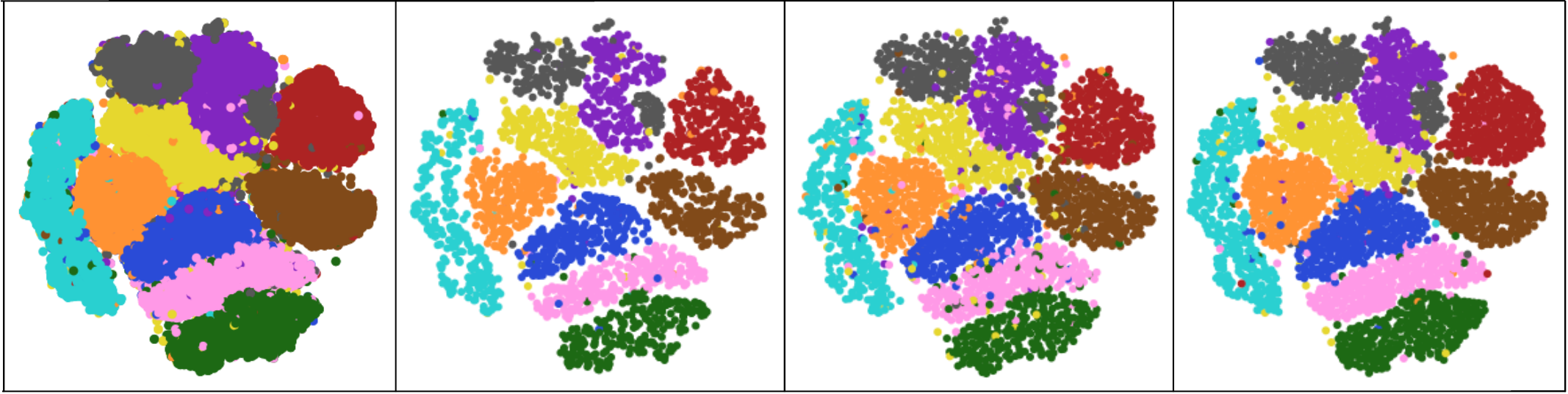}
  \put(0.5, 1){(a)}
  \put(25.5, 1){(b)}
  \put(50.5, 1){(c)}
  \put(75.5, 1){(d)}
  \end{overpic}
  \caption{Different sampling results of the MNIST dataset: (a) the original scatterplot; (b) the result of random sampling; (c) the result of outlier biased density based sampling; (d) the result of blue noise sampling. The three sampling methods obtain satisfying performances in preserving relative region density, outliers and the overall shapes in terms of human perception, respectively.
}
\label{fig:teaser}
}

\abstract{
Given a scatterplot with tens of thousands of points or even more, a natural question is which sampling method should be used to create a small but "good" scatterplot for a better abstraction. 
We present the results of a user study that investigates the influence of different sampling strategies on multi-class scatterplots. 
The main goal of this study is to understand the capability of sampling methods in preserving the density, outliers, and overall shape of a scatterplot.
To this end, we comprehensively review the literature and select seven typical sampling strategies as well as eight representative datasets.
We then design four experiments to understand the performance of different strategies in maintaining: 1) \yuanjun{region} density; 2) \yuanjun{class density}; 3) outliers; and 4) overall shape in the sampling results. 
The results show that: 1) \yuanjun{random sampling is} preferred for preserving region density; 2) blue noise sampling and random sampling have comparable performance \yuanjun{with the three multi-class sampling strategies} in preserving class density; 3) \yuanjun{outlier biased density based sampling, recursive subdivision based sampling, and blue noise sampling} perform the best in keeping outliers; and 4) \yuanjun{blue noise sampling} outperforms the others in maintaining the overall shape of a scatterplot. 
}


\begin{document}




\fontsize{9}{9} 

\firstsection{Introduction}
\maketitle
Scatterplots are one of the most widely used \shixia{visual representations in exploratory data analysis}~\cite{micallef2017towards,Sarikaya18}.
Their flexibility enables \doc{the discovery of} free-form patterns in \doc{two-dimensional} data, such as trends, clusters, and outliers~\cite{Kandogan12}.
\doc{In conjunction with} dimensionality reduction approaches, scatterplots are also the dominant visualization tool \doc{for exploring} high-dimensional data~\cite{liu2019crowsourcing,Sacha17,sedlmair2013empirical}.
However, scatterplots become less effective when data grows in \lingyun{size}. 
First, the overdraw issue will \doc{adversely impact} the \doc{ability to comprehend} scatterplots~\cite{Mayorga13}.
Second, the speed of \lingyun{producing visualization}, \ie, loading and rendering source data, will \doc{become} a considerable issue~\cite{Park2015}.

\looseness=-1
\revision{
Many efforts have been devoted to addressing the overplotting issues in scatterplots, including sampling~\cite{Dix2002, Ellis2005}, abstraction~\cite{liu2017towards,Zhao2019phoenixmap}, modifying the size~\cite{Li2010size, Woodruff1998} and the opacity~\cite{Li2010lightness, Mayorga13} of the visual marks, and other hybrid methods~\cite{Matejka2015}.
However, many of them still suffer from the scalability problem~\cite{Ellis2007} since they still need to render a large number of data points or execute complex computations to produce the visualization, which restricts their practical use due to limited visualization capability of the display devices or computational resources.}
To overcome the scalability issue in scatterplots, sampling has been well studied in data mining~\cite{Park2015} and visualization~\cite{Bertini2006, Chen14,yuan2021survey}.
Generally, sampling aims to select a statistically unbiased representation of the full dataset.
In different scenarios, many sampling strategies have been developed to enhance specific aspects of the full dataset, \eg, density~\cite{Bertini2006}, outlier~\cite{Xiang19}, \shixia{shape~\cite{Liu2017}}, and class ratio~\cite{Chen14, Chen20, Hu20}.
As shown in Fig.~\ref{fig:teaser}, the three sampling strategies preserve different features in their sampling results.
Rojas~\etal~\cite{Rojas17} interviewed 22 data scientists and concluded that random sampling, which is statistically unbiased, is the only \doc{appropriate} choice \doc{open to} these scientists for data exploration.
Although other sampling strategies can provide different insights for data exploration, data scientists are not familiar with them, and \lingyun{they} do not know which strategy to use in a specific scenario.
For instance, \lingyun{although} \yuanjun{blue noise sampling} \lingyun{has been} widely used in computer graphics and visualization, we have not found \doc{examples of} its application in data mining in our literature review.

\looseness=-1
Nevertheless, the researches \doc{into} sampling strategy design have \doc{conducted} many quality comparisons.
\doc{However,} performing a perception-based evaluation study is still essential \doc{for providing} guidelines \doc{for} choosing sampling strategies.
On the one hand, most of the existing comparisons are based on objective quality measures, \eg, density, class ratio, and \yuanjun{the} number of outliers.
Their conclusions may not be suitable for visualization tasks due to perceptual biases~\cite{Ma18, Wei20}.
On the other hand, these strategy-oriented evaluations are limited to a subset of tasks and approaches.
\shixia{Thus, a} comprehensive evaluation \shixia{of representative approaches} is missing.

In this paper, we conduct four experiments to study the effects of typical sampling strategies on 2D scatterplots.
First, we select seven strategies based on a comprehensive literature review.
They are either widely used or task-specific sampling strategies that have specific goals.
The strategies include random sampling~\cite{mukhopadhyay2008theory}, blue noise sampling~\cite{Cook1986}, density biased sampling~\cite{palmer2000density}, multi-class blue noise sampling~\cite{Chen14, wei2010multi}, outlier biased density based sampling~\cite{Xiang19}, multi-view Z-order sampling~\cite{Hu20}, and recursive subdivision based sampling~\cite{Chen20}.
Second, we identify four typical analytical tasks in multi-class \doc{scatterplot} analysis, including identifying relative \yuanjun{region} density, relative \yuanjun{class} density, outliers, and shapes.
Third, we formulate \yuanjun{four} hypotheses based on our experience and literature review.
We hypothesize that
(1) for a scatterplot without class information, all other sampling strategies perform better than \shixia{random sampling} in relative \yuanjun{region} density identification tasks in terms of accuracy and efficiency;
\yuanjun{(2) for a scatterplot with class information, multi-class sampling strategies perform better than other sampling strategies in relative class density identification tasks in terms of accuracy and efficiency;}
(3) \shixia{outlier biased density based sampling is the best} in the outlier identification task;
(4) \shixia{blue noise sampling \yuanjun{and multi-class blue noise sampling} perform better than other strategies in preserving the overall shape.}

We select eight datasets that present different patterns and \lingyun{various degrees of visual clutter}.
100 participants are recruited for the \lingyun{formal} experiments.
Before the formal study, we perform a pre-study \lingyun{\doc{with} \yuanjun{160} participants} to determine the sampling ratio and color stimuli.
In the formal study, \lingyun{we conduct a series of experiments on different sampling strategies and datasets.}
We also design subjective questionnaires to \lingyun{obtain} the subjective experience of the participants.

Based on the experiment results, we perform a comprehensive statistical analysis.
\yuanjun{The analysis results of the objective metrics} suggest that (1) \shixia{H1 is rejected;} with random sampling, participants use \doc{less} time to complete the \yuanjun{region} density identification tasks \jiazhi{with higher accuracy;}
(2) H2 is partially confirmed; \jiazhi{multi-class sampling strategies \doc{achieved} higher accuracy than other strategies except for blue noise sampling; with random sampling, participants use \doc{less} time to complete the class density identification tasks.}
(3) H3 is \shixia{partially} confirmed; outlier biased density based sampling, recursive subdivision based sampling, and blue noise sampling \jiazhi{perform better than other strategies in identifying outliers.}
(4) H4 is \yuanjun{partially} confirmed; \jiazhi{blue noise sampling performs the best in shape \doc{preservation} while multi-class blue noise sampling performs at a middle level.}
The analysis results of the subjective questionnaires provide useful insights into the sampling strategies.
\lingyun{They disclose} subjective reasons for the objective metric results.
After the analysis, we summarize the ability of the seven sampling strategies to support our identified tasks.

In summary, we present a comprehensive perception-based evaluation of sampling strategies for scatterplots.
We contribute a carefully designed evaluation and a series of instructive findings, \doc{offering} guidelines for choosing sampling strategies in task-specific scenarios.
\shixia{In addition, we also contribute a Python library for scatterplot sampling, which contains \yuanjun{14} commonly used sampling \lingyun{algorithms} and is available at} \url{https://github.com/libsampling/libsampling}.

\section{Related Work}
\label{sec:related-work}

\subsection{Sampling Strategies for Scatterplots}
\label{sec:categories}
\shixia{The scatterplot sampling methods can be categorized into two classes: single-class sampling and multi-class sampling.}

\vspace{1mm}
\noindent\textbf{Single-class sampling.}
\shixia{This category of sampling strategies aims to preserve the properties of interest (\eg, density) of the original dataset without considering class information.}
Random sampling, the most widely used sampling method, is a classical single-class sampling method.
\shixia{It employs a \revision{\textbf{uniform sampling}} strategy that treats all samples equally and selects each sample with the same probability.}

\looseness=-1
On the contrary, \revision{\textbf{non-uniform sampling}} strategies assign varying sampling probability to \lingyun{data} so that \lingyun{some specific properties of the original datasets can be better preserved.}
For example, in some cases, samples are required to be 
\jiazhi{better} spatially separated~\cite{Liu2017, zhang2014visual}.
Blue noise sampling~\cite{zhang2014visual, yan2015survey} achieves this by selecting samples with blue noise properties so that the selected samples will distribute 
evenly in the sample space.
Farthest point sampling~\cite{berger2016cite2vec} can also select samples with better \yuanjun{spatial} separation.
It randomly picks the first sample, and then \lingyun{iteratively} selects samples of maximal minimum distances to the previously selected ones.
Liu~\etal~\cite{Liu2017} developed a \jiazhi{dual space} sampling strategy. 
It computes a density field of the original sample space and maps the samples from the original density space to a uniform density space through a warping function.
Then it selects the samples via orthogonal least squares or weight sample elimination \jiazhi{in the mapped space} in order to maintain good spatial separation among the selected samples.
Lastly, the selected samples are mapped \jiazhi{back} into the original density space.

There are also sampling strategies \doc{that have been} developed to preserve density-related properties.
Density biased sampling~\cite{palmer2000density} tends to over-sample sparse regions and under-sample dense regions in the sample space.
It can counterbalance samples from both regions, thus preserving small clusters and more solitary samples.
Bertini~\etal~\cite{bertini2004chance, Bertini2006} proposed a non-uniform sampling strategy aiming at preserving the relative region density difference.
It divides the sample space into uniform grids, and then determines the represented density of each grid and finally selects samples from each grid according to \lingyun{the density}.
Joia~\etal~\cite{Joia2015} formulated the sampling problem as a matrix decomposition problem and solved it with singular value decomposition (SVD).
\lingyun{This method} performs SVD on the original dataset and selects the samples with the biggest correlation with top-$k$ basis vectors in the SVD result, where $k$ is a rank parameter indicating the number of principal components of interest.
The SVD based sampling strategy can counterbalance the number of points from regions with different densities.

Outlier preservation is another common goal in sampling strategies.
A typical method \lingyun{for achieving this goal} is to alter \lingyun{existing} sampling strategies, making them probabilistically accept more outliers according to specified outlier scores~\cite{liu2017visual, Xiang19}.
For instance, Liu~\etal~\cite{liu2017visual} proposed outlier biased random sampling that assigns higher sampling probabilities to outliers in random sampling.
Xiang~\etal~\cite{Xiang19} increased the accepting probability of outliers in the sampling process of blue noise sampling and density biased sampling and developed outlier biased blue noise sampling and outlier biased density based sampling, respectively.
\lingyun{Moreover}, Cheng~\etal~\cite{Cheng2019} sampled the point clouds on their color mapping display using a hashmap based stratified sampling technique to preserve outliers while keeping the main distribution.

\begin{table}[t]
\vspace{1mm}
\caption{Characteristics of our collected sampling strategies. MC refers to multi-class sampling strategies; NU refers to non-uniform sampling strategies; S refers to considering spatial separation; D refers to considering density; O refers to considering outlier preservation. \revision{The selected sampling strategies are bold, and their acronyms are attached.}}
\vspace{-2mm}
\resizebox{0.49\textwidth}{!}{
\small
\renewcommand\arraystretch{1.2}
\centering
\revision{
\begin{tabular}{|l|l|c|c|c|c|c|}
\hline
Sampling strategy & Works & MC & NU & S & D & O \\
\hline
\textbf{Random sampling (RS)} & \makecell[l]{\cite{dosSantosAmorim2012}, \cite{Liu2016Grassmannian}, \cite{Poco2011}, \cite{Rieck2015}, \\ \cite{Tao2019}, \cite{Xia2018}, \cite{Zhao2020}}  &  &  &  &  &  \\
\hline
\textbf{Blue noise sampling (BNS)} & \cite{Chen14}, \cite{Xiang19}, \cite{zhang2014visual}  & & $\surd$ & $\surd$ & & \\
\hline
Farthest point sampling & \cite{berger2016cite2vec}  & & $\surd$ & $\surd$ & & \\
\hline
Dual space sampling & \cite{Liu2017}  & & $\surd$ & $\surd$ & & \\
\hline
\textbf{Density biased sampling (DBS)} & \cite{Xiang19}  & & $\surd$ & & $\surd$ & \\
\hline
Non-uniform sampling & \cite{bertini2004chance}, \cite{Bertini2006}  & & $\surd$ & & $\surd$ & \\
\hline
SVD based sampling & \cite{Joia2015}  & & $\surd$ & & $\surd$ & \\
\hline
Outlier biased random sampling & \cite{liu2017visual}, \cite{Zhao2019}  & & $\surd$ & & & $\surd$ \\
\hline
\textbf{Outlier biased density based sampling (OBDBS)} & \cite{Xiang19}  & & $\surd$ & & $\surd$ & $\surd$ \\
\hline
Outlier biased blue noise sampling & \cite{Xiang19}  & & $\surd$ & $\surd$ & & $\surd$ \\
\hline
Hashmap based sampling & \cite{Cheng2019}  & & $\surd$ & & & $\surd$ \\
\hline
\textbf{Multi-class blue noise sampling (MCBNS)} & \cite{Chen14}  & $\surd$ & $\surd$ & $\surd$ & & \\
\hline
\textbf{Multi-view Z-order sampling (MVZS)} & \cite{Hu20}  & $\surd$ & $\surd$ & & $\surd$ & \\
\hline
\textbf{Recursive subdivision based sampling (RSBS)} & \cite{Chen20}  & $\surd$ & $\surd$ &  & $\surd$ & $\surd$ \\
\hline
\end{tabular}
}
}
\label{tab:category}
\vspace{-7mm}
\end{table}

\begin{figure*}[t]
\centering
\includegraphics[width=\linewidth]{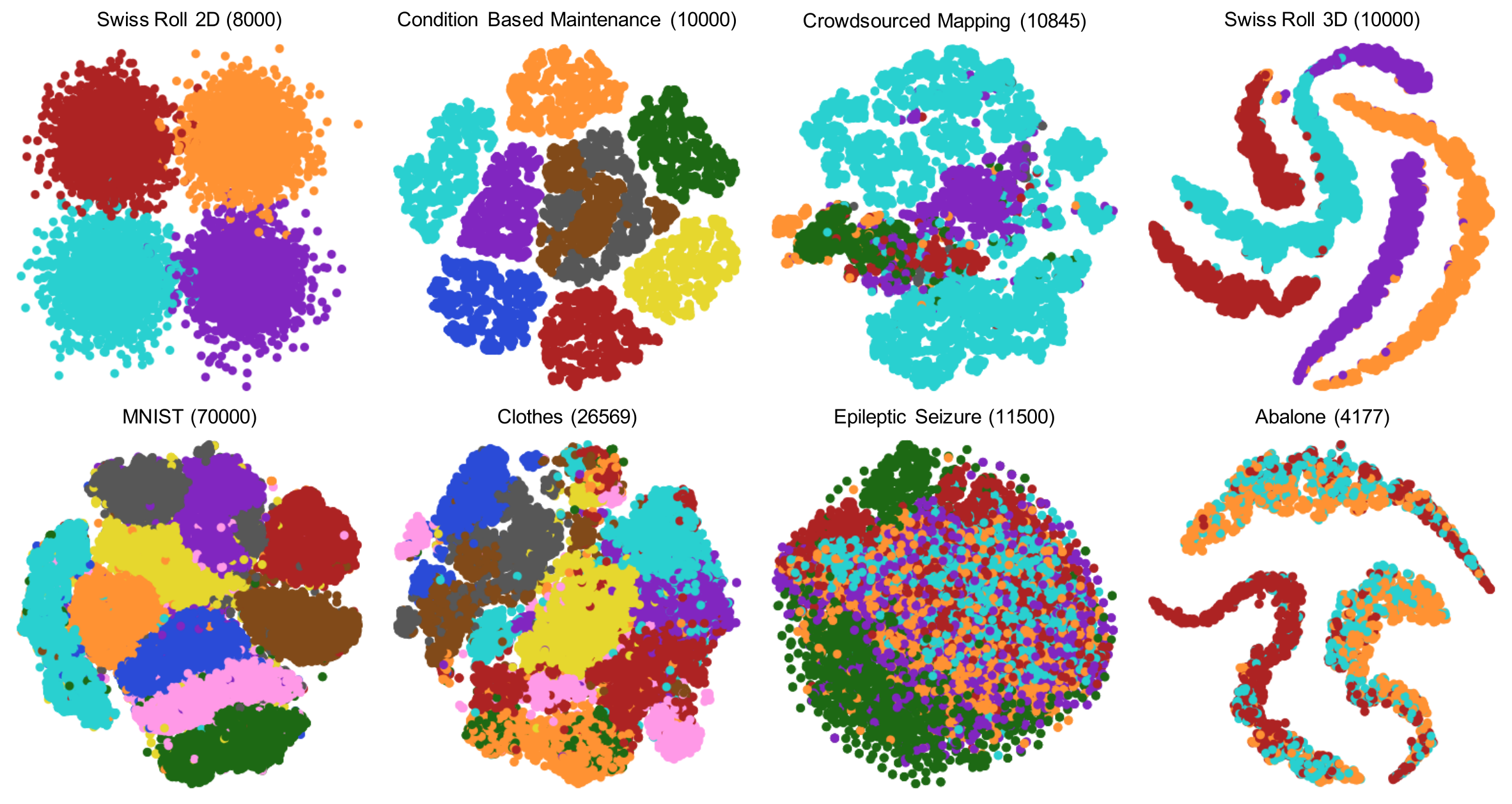}
\vspace{-5mm}
\caption{Datasets selected for our evaluation. \revision{All datasets are multi-class data. The color encodes the labels.} The numbers in the brackets indicate the sizes of the datasets.}
\vspace{-5mm}
\label{fig:datasets}
\end{figure*}

\vspace{1mm}
\noindent\textbf{Multi-class sampling.}
Unlike single-class sampling strategies, multi-class sampling strategies aim to preserve the properties of interest (\eg, density) of each individual class as well as their union.
\revision{Thus, all the multi-class sampling methods are \textbf{non-uniform}.}
Wei~\cite{wei2010multi} extended blue noise sampling to multi-class scenarios to maintain the blue noise properties of \lingyun{each class of samples and of the whole dataset}.
Based on the multi-class blue noise sampling, Chen~\etal~\cite{Chen14} employed a hierarchical sampling strategy that selects samples round by round.
It first selects samples from the coarsest level using multi-class blue noise sampling, and when the selected samples are not enough, it reduces the restricted distance of the selected samples by half and adds more samples in the final result.
Recently, a recursive subdivision based sampling strategy proposed by Chen~\etal~\cite{Chen20} \doc{met} several requirements \doc{for} multi-class scatterplot exploration, including preserving relative densities, maintaining outliers, and minimizing visual artifacts.
It splits the visual space into a binary KD-tree and determines which class of instances should be selected at each leaf node based on relative class density by a backtracking procedure.
Additionally, Hu~\etal~\cite{Hu20} developed multi-view Z-order sampling based on Z-order curve methods~\cite{Zheng2013} and formulated it as a set cover problem.
The sets were constructed by segmenting the Z-order curves of the samples in each class and the whole dataset, respectively.
\lingyun{This strategy} selects samples by greedily solving such set cover problems and gets satisfying results in terms of minimizing kernel density estimation error.

\subsection{Evaluation Studies of Sampling Methods}



A few \doc{previous} studies paid attention to the evaluation of sampling methods.
\lingyun{However, they} either \doc{evaluated} the sampling methods in certain situations (\eg, graph sampling) or \doc{evaluated} a specific sampling method to show its capability.

\vspace{1mm}
\looseness=-1
\noindent\textbf{Generic Evaluation}.
Previous studies concentrated on evaluating sampling methods for graph data~\cite{wu2016evaluation, nguyen2017proxy}.
Wu~\etal~\cite{wu2016evaluation} conducted a survey on graph sampling methods and performed an empirical evaluation \doc{of} the preservation of the three most important visual factors on five selected methods.
Later, Nguyen~\etal~\cite{nguyen2017proxy} proposed a family of quality metrics to evaluate the stochastic graph sampling methods in an objective manner.

\vspace{1mm}
\noindent\textbf{Instance-oriented Evaluation}.
When proposing new sampling methods, researchers \doc{also conduct} evaluations to demonstrate their effectiveness.
Some of them used quality measures to make a quantitative evaluation in terms of data features.
Chen~\etal~\cite{Chen20} adopted four measures based on their design requirements and compared their results with three baseline methods.
They also presented three case studies to show the usefulness of their method in multi-dimension data analysis.
However, as numerical measures \lingyun{do not always agree with} human perception~\cite{wang2019improving}, other efforts focused on empirically evaluating perceptual subjects through user studies.
For example, both hierarchical multi-class blue noise sampling~\cite{Chen14} and multi-view Z-order sampling~\cite{Hu20} employed user studies  to confirm their superiority \doc{in} the recognition of data classes and densities.

\looseness=-1
To the best of our knowledge, there has never been a systematic evaluation of sampling on scatterplots from the perspective of visualization.
In this paper, we collect the representative sampling strategies on scatterplots from the visualization community and then conduct four experiments to evaluate their ability to retain data features on perception.

\section{Evaluation Landscape}
\label{sec:Hypothesis}

\looseness=-1

\subsection{Selection of Sampling Strategies}
\label{sec:SelectionOfSamplingMethods}

To comprehensively summarize the sampling strategies used in the visualization community, we first surveyed papers from 
\jiazhi{the journal of IEEE TVCG and three mainstream visualization conferences (IEEE VIS, PacificVis, and EuroVis)}
published from 2010 to 2019.
We used Google Scholar to search for the papers with the keyword "sampling" from the \doc{above sources}.
There are 1,562 papers in the initial result.
\lingyun{Next, we filtered out papers \doc{that} are not relevant to sampling in visualization.}
\lingyun{We kept papers that either applied sampling for visualization purposes or proposed new sampling strategies in visual analytics or information visualization.}
Finally, \lingyun{\yuanjun{25} papers remained in our survey. We further summarized the sampling strategies discussed in these papers}.

The collection of the sampling strategies are \yuanjun{listed} in Table~\ref{tab:category}.
\lingyun{We decided to focus on the widely used or} 
\jiazhi{recently advanced} \yuanjun{task-specific}
\lingyun{sampling strategies since there are diverse sampling strategies used for different visualization purposes, and it is obviously \yuanjun{impractical} to evaluate all \doc{of them} in our work.}
\revision{As a result, we selected sampling strategies that cover all four categories in two dimensions, single-class/multi-class sampling and uniform/non-uniform sampling.}
We first selected random sampling (RS)~\cite{mukhopadhyay2008theory}, because it is the most widely used sampling strategy.
We also selected other representative strategies, including blue noise sampling (BNS)~\cite{Cook1986}, density biased sampling (DBS)~\cite{palmer2000density}, and multi-class blue noise sampling (MCBNS)~\cite{Chen14, wei2010multi}.
In addition, outlier biased density based sampling (OBDBS)~\cite{Xiang19}, multi-view Z-order sampling (MVZS)~\cite{Hu20}, and recursive subdivision based sampling (RSBS)~\cite{Chen20} are selected, \revision{since they are reported to} perform the best in terms of their design requirements \revision{based on the experiment results from the previous studies}, respectively.
\revision{
For instance, outlier biased blue noise sampling outperforms four other sampling methods in terms of preserving outliers and class consistency in the experiment~\cite{Xiang19}.}
These seven strategies \doc{have} all \doc{been} \yuanjun{included} in our study.

\subsection{Selection of Datasets}
\looseness=-1
\lingyun{To ensure the reliability of the evaluation results,}
we selected datasets from the previous studies in visualization as our experiment data.
\lingyun{More specifically, we collected datasets that were used in the works in our survey.}
\lingyun{Since} most of them are high-dimensional data,
\lingyun{we first transformed them into 2D \yuanjun{data} using t-SNE and normalized \yuanjun{them} to $[0, 1] \times [0, 1]$.}
\lingyun{In the results, points} are located 
\jiazhi{as clusters in different shapes}
in the obtained multi-class scatterplots.
In addition, the number of points and the clutter degrees of these scatterplots vary within a wide range.
According to the observations above, we selected eight representative datasets \yuanjun{shown in Fig.~\ref{fig:datasets}} with different characteristics: \lingyun{six datasets} where points are located as clusters (\emph{Swiss Roll 2D}~\cite{SwissRoll04}, \emph{Condition Based Maintenance}~\cite{Coraddu2014}, \emph{Crowdsourced Mapping}~\cite{Johnson2016}, \emph{MNIST}~\cite{Lecun1998}, \emph{Clothes}~\cite{Xiang19}, and \emph{Epileptic Seizure}~\cite{Andrzejak2001}); and two where points are located as 
\jiazhi{curved stripes}
(\emph{Swiss Roll 3D}~\cite{SwissRoll04} and \emph{Abalone}~\cite{Dua2019}).
The clutter degrees of them vary from slight to severe as the order listed in each bracket.
\lingyun{The number of points in the selected datasets} ranges from thousands to tens of thousands, \yuanjun{as listed in Fig.~\ref{fig:datasets}}.


\subsection{Selection of Visual Factors}
\label{sec:factor}
\looseness=-1
We identified the most critical visual factors for the sampling strategies by comprehensively reviewing existing works on sampling and scatterplots.
\shixia{Previous studies have shown \lingyun{that} there are basically five goals related to scatterplot exploration, \yuanjun{including} outlier identification, shape examination, trend analysis, density detection, and coherence analysis~\cite{micallef2017towards,Wilkinson2005}.}
\yuanjun{In order to determine which of these factors mentioned in the aforementioned goals are \shixia{concerned in the existing sampling strategies},}
\shixia{we carefully \doc{examined} the \yuanjun{25} selected papers and extracted the visual factors that are considered in these works.}
\shixia{Specifically,}
outlier maintenance is mentioned \doc{most often}, \doc{as it} appears in seven \yuanjun{out of the 25 papers}.
\lingyun{Three papers concern preserving} the relative density and \lingyun{two concern} the overall shape of a scatterplot, respectively.
\revision{Here, \emph{the overall shape} refers to the geometric properties of the distribution of samples on a scatterplot, \eg, whether a set of scatter points are convex or not.}
As a result, in the study, we decided to investigate the capabilities of different sampling strategies in preserving \textbf{outliers}, \textbf{density}, and the \textbf{overall shape} of a scatterplot.

\section{Pre-Study}
\looseness=-1
The purpose of the pre-study is to specify two experiment choices for the formal study: (1) how many points should be sampled from each dataset, and (2) 
\jiazhi{whether color encoding should be used in the experiment \doc{for} comparing region densities of multi-class scatterplots in the formal study.}
\jiazhi{\emph{Region density} refers to the density of data points regardless of class information.}

\begin{figure}[t]
\centering
\includegraphics[width=\linewidth]{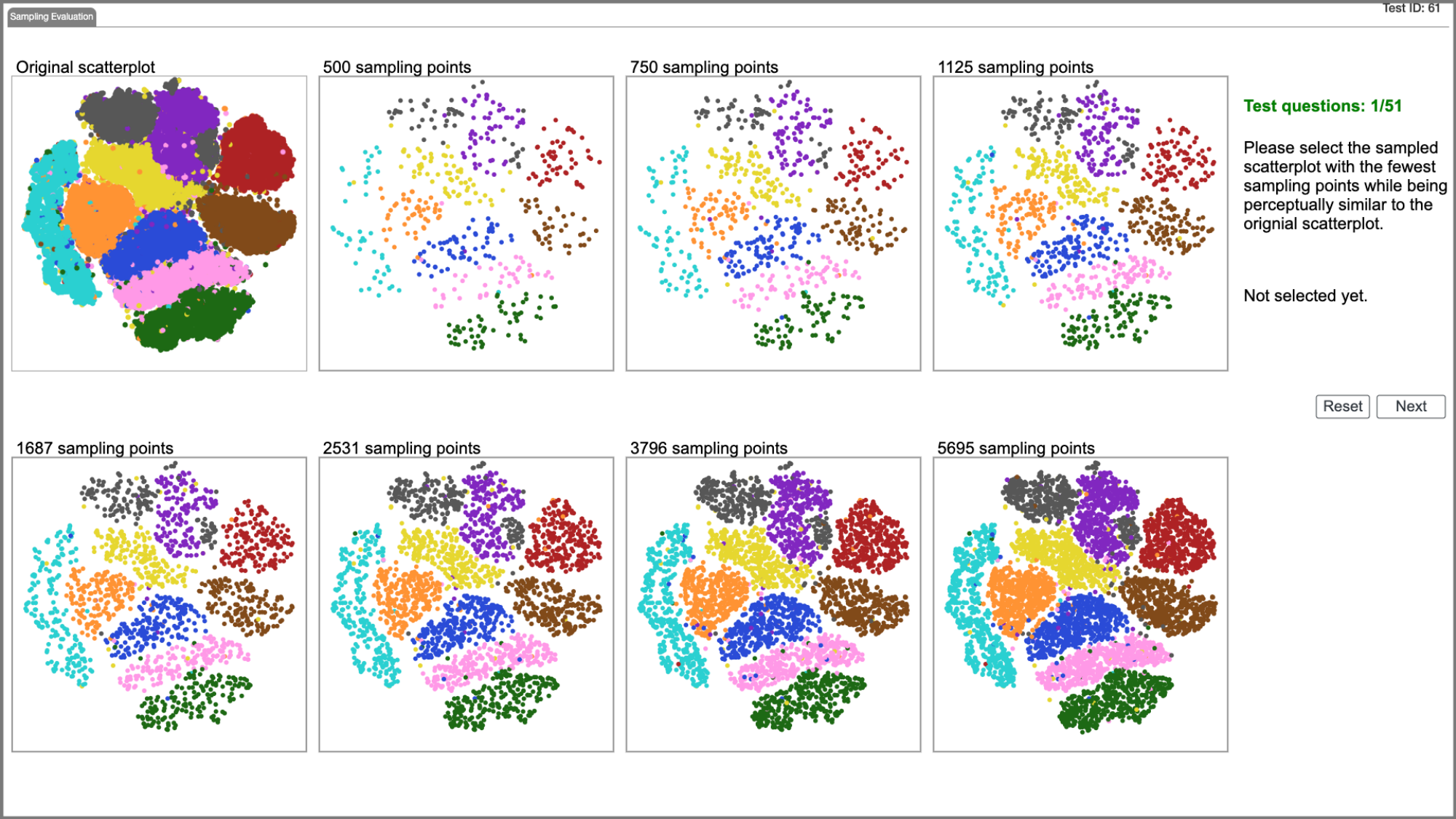}
\vspace{-5mm}
\caption{Interface of Experiment 1 in the pre-study: the original scatterplot is located at the top-left, with the sampling results of increasing sampling number at the following positions.}
\vspace{-6mm}
\label{fig:interface_1}
\end{figure}

\subsection{Experiment 1: Sampling Number Identification}
\looseness=-1
We conducted a subjective experiment to specify the proper number of sampling points for each dataset.
On the one hand, we want set a small sampling number to clearly show the motivation of sampling, \yuanjun{\ie}, addressing the issue of visual clutter.
On the other hand, the patterns of the original scatterplots will not be preserved if the sampling number is too small.
Because the patterns are different in different datasets, it is essential to choose a proper sampling number for each dataset.



\vspace{1mm}
\noindent\textbf{Task and Procedure}.
We used the same \jiazhi{eight} datasets as those in the formal study.
For each dataset, we showed the participants the original scatterplot as well as a series of sampled scatterplots with different sampling numbers.
Participants were asked to select \lingyun{the sampled scatterplot that has the smallest number of points while being perceptually similar to the original} \jiazhi{scatterplot.} 
Weber-Fechner Law states that the perceived intensity is proportional to the logarithm of the stimulus~\cite{portugal2011weber}. 
Therefore, we adopted a seven-level sampling with $500$, $500 \times 1.5$, $500 \times 1.5^2$, ..., $500 \times 1.5^6$ points (a geometric sequence) by \lingyun{random sampling}.
\jiazhi{We \doc{chose} random sampling because it has no preference for \revision {preserving} certain \doc{properties} of \doc{the} datasets.}
\revision{In addition, random sampling will not introduce bias to the sampled data statistically and guarantee the representativeness of the sampled data~\cite{verma2019statistics}.}
\jiazhi{Leskovec~\etal~\cite{Leskovec2006} show that different sampling \yuanjun{strategies} produce very similar results when the sampling rate is greater than 50\%.}
\jiazhi{Therefore, we cut off the sampled scatterplots in a sequence if their sampling rate \doc{was} greater than 50\% to avoid meaningless comparison.}
As a result, there \doc{were} seven sampled scatterplots \lingyun{(at most), along with the original scatterplot with all points}, to be displayed.
\lingyun{The eight scatterplots \doc{were}} arranged in a $2 \times 4$ matrix layout, as shown in Fig.~\ref{fig:interface_1}.
\revision{There were eight trials for each participant in this experiment, corresponding to eight datasets, respectively.}
It took about 5 minutes for each participant to finish the experiment.
\yuanjun{Participants were asked which visual factors they \doc{were concerned with} in the post-experiment questionnaire.}

\begin{figure*}[ht]
\centering
\vspace{-5mm}
\includegraphics[width=\linewidth]{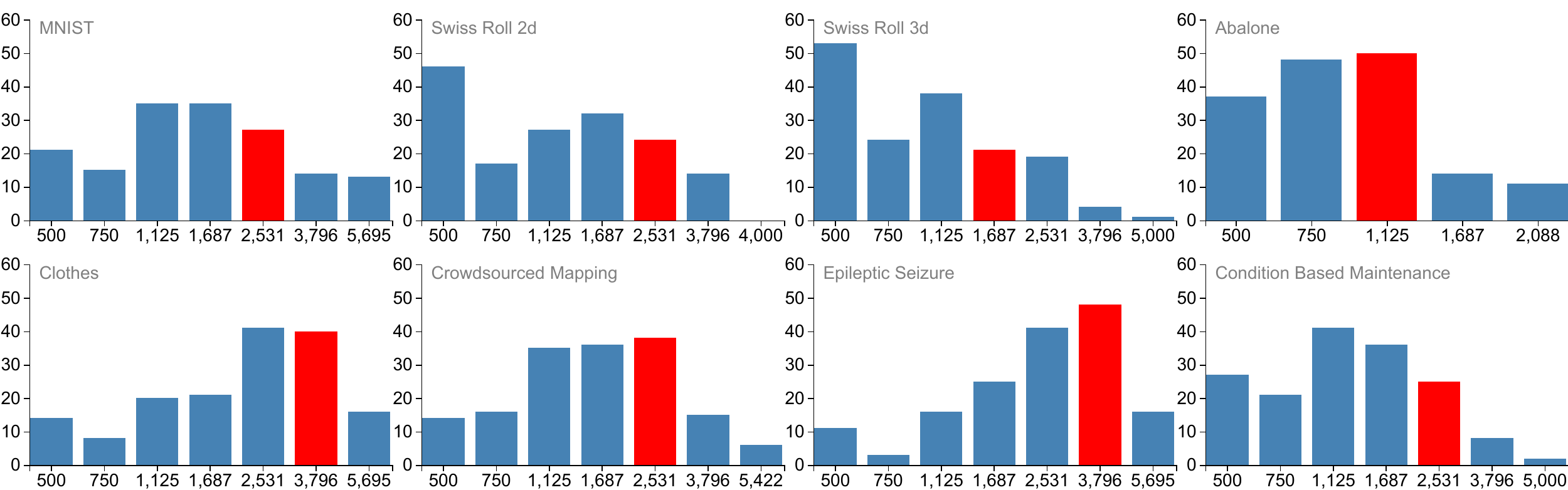}
\vspace{-5mm}
\caption{Results of Experiment 1 in the pre-study (Sampling number identification). This bar chart shows the number of votes for each sampling number in each dataset. Red bars indicate the selected sampling numbers in our formal study.
}
\vspace{-3mm}
\label{fig:pilot_task1}
\end{figure*}

\begin{figure*}[t]
\centering
\includegraphics[width=\linewidth]{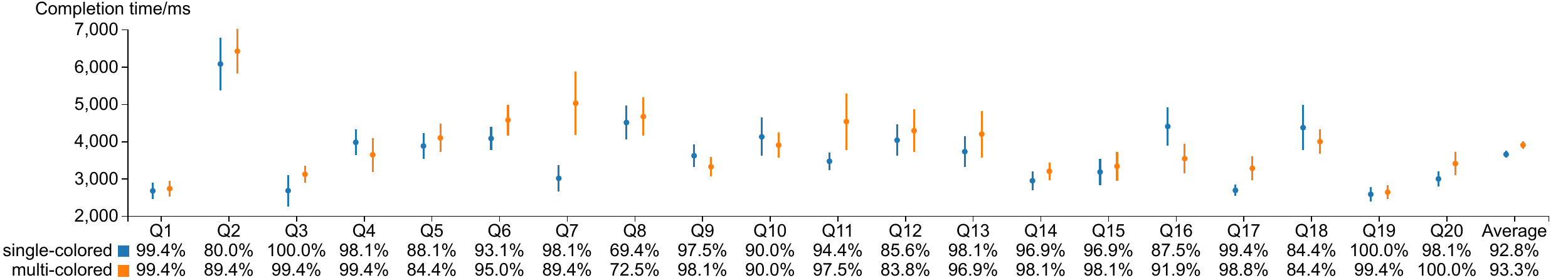}
\vspace{-5mm}
\caption{Results of Experiment 2 in the pre-study (Understanding color effect on region density identification). Q1 -- Q20 stands for the 20 generated questions. Error bars indicate 95\% confidence intervals \revision{(same as below)}. Below the x-axis listed the accuracy of each question.}
\vspace{-5mm}
\label{fig:pilot_task2}
\end{figure*}

\vspace{1mm}
\noindent\textbf{Participants and Apparatus}.
We recruited \yuanjun{160} participants (\yuanjun{130} males, \yuanjun{30} females, aged 18 -- 60 years).
All \doc{the} participants were either students or researchers with \doc{a} computer science background.
69 participants reported \doc{being} familiar or very familiar with visualization, 32 moderately familiar, and 61 unfamiliar or very unfamiliar.
65 participants reported previous experience with sampling.

\looseness=-1
The experiment was conducted through a web prototype.
Participants were asked to perform the experiment on a screen with a resolution higher than $1,920 \times 1,080$.
\revision{
The points in the scatterplots were rendered with a radius of 3 pixels, which was preliminarily confirmed by iterative adjustments within a common range (1--5 pixels).
Moreover, we rendered the points without transparency to avoid affecting the color perception~\cite{d1997color}.
}

\vspace{1mm}
\looseness=-1
\noindent\textbf{Results}.
Given the fact that when there are more points in \doc{a} sampled scatterplot, it will look more like the original scatterplot, \doc{we assumed that participants would consider the sampling results still similar to the original scatterplot when the sampling number was more than the selected one.}
\jiazhi{In addition, considering that other sampling strategies may perform better than random sampling, we needed to leave the space to show their superiority in our experiments.}
\lingyun{Based on these considerations, we have chosen the optimal sampling number by requiring that the sampling numbers of the scatterplots selected by 80\%, rather than 100\%, of the participants were smaller than the optimal one.}
\yuanjun{Fig.~\ref{fig:pilot_task1} presents the \lingyun{results, which shows that the optimal} sampling number for most datasets (\emph{MNIST}, \emph{Swiss Roll 2D}, \emph{Crowdsourced Mapping}, and \emph{Condition Based Maintenance}) is 2,531, while the sampling rates of them were $3.6\%$, $31.6\%$, $23.3\%$, and $25.3\%$, respectively.}
\yuanjun{Four exceptions were the datasets \emph{Clothes}, \emph{Epileptic Seizure}, \emph{Swiss Roll 3D}, and \emph{Abalone}, whose optimal sampling numbers with the corresponding sampling rates were 3,796 ($14.3\%$), 3,796 ($33.0\%$), 1,687 ($16.9\%$), and 1,125 ($26.9\%$), respectively.}
\jiazhi{According to the results of the subjective questionnaire in the experiment,}
\lingyun{when judging the similarity between scatterplots, over 75\% of the participants took the overall shape of each class of points into consideration, followed by density 
($55\%$) and outliers ($35\%$).}

\subsection{Experiment 2: \yuanjun{Understanding Color Effect on Region Density Identification}}

This \jiazhi{controlled} experiment aims \doc{to understand} the \lingyun{effect} of color when comparing \jiazhi{region} density \jiazhi{in multi-class scatterplots}.
\jiazhi{Though color is not related to the definition of region density, encoding class labels with color may affect human perception.}
\revision{Therefore, we should figure out whether color affects the perception of region density to decide whether color to be used in the formal study.}


\noindent\textbf{Task and Experiment design}.
\yuanjun{We generated ten synthetic datasets using mixed Gaussian distributions with 3 to 10 classes.}
\jiazhi{These datasets} were different from the ones in the formal study.
\revision{In each question, two rectangular regions were marked on the same scatterplot, and participants were asked to select the region with the higher density.}
This experiment adopted a within-subject design, and the only variable was whether the scatterplot was colored or not.
\lingyun{We asked the same question\yuanjun{s} \yuanjun{with multiple colors or with only a dark grey color}.
\lingyun{We provided two questions} on each synthetic dataset, \yuanjun{so there were 20 different questions.}
Thus, in total, we had}
$$
\setlength{\abovedisplayskip}{3pt}
\setlength{\belowdisplayskip}{3pt}
    \yuanjun{160\ (participants)} \times 20\ (questions) \times 2\ (colors) = 6,400
$$
results.
In order to eliminate the learning effect, \yuanjun{the multi-colored and single-colored versions} of the same question were arranged \lingyun{to appear in random order and not consecutively}.
It took about 5 minutes for each participant to finish the experiment.


\looseness=-1
\noindent\textbf{Procedure}.
\lingyun{In order to help the participants \doc{become} familiar with the task, the experiment started with a training session of three questions.
In the training session, the correct answers were shown to the participants after they submitted their answers.
Participants could ask questions during the training session.
Time and accuracy were not recorded.
As long as the participants reported that they had fully understood the experiment, we started the real \yuanjun{test}, where completion time and accuracy for each question were recorded.
Thus, in the real study, we reminded participants that they needed to finish the experiment as fast and precisely as possible.
After the experiment, participants were asked to finish a questionnaire and rate the color \lingyun{effect on} a five-point Likert scale.
}


\noindent\textbf{Participants and Apparatus}.
\lingyun{We had the same participants and used the same apparatus as in Experiment 1.}

\noindent\textbf{Results}.
\revision{As Fig.~\ref{fig:pilot_task2} shows,}
the average accuracy of the \yuanjun{multi-colored} and \yuanjun{single-colored} questions were 92.9\% and 93.3\%, respectively, while the average completion time of the \yuanjun{multi-colored} questions was \lingyun{3,581ms and \yuanjun{that} of the \yuanjun{single-colored} ones was 3,616ms}.
\yuanjun{Since the data was not subject to the normal distribution according to the Shapiro-Wilk test, we conducted \yuanjun{the} Wilcoxon test to \yuanjun{examine} the significance of their difference with \doc{a} significance level \doc{of} $\alpha = 0.05$.
No statistical significance in the difference of their accuracy is reported through \lingyun{hypothesis} tests ($p = 0.2648 > 0.05$).
But \doc{a} significant difference \doc{was shown} in terms of completion time ($p = 1.551 \times 10^{-10} < 0.001$), \revision{indicating} that participants spent significantly \doc{less} time in completing \yuanjun{single-colored} questions than multi-colored ones.}
\revision{The subject feedback reflected that participants felt that color affected the region density comparison slightly with an average score of 2.03 of the color deficiency.}
This is also \doc{consistent} with our numerical results.
A participant commented that \emph{"the salient color may have an influence on my judgment of density."}
Consequently, we decided to \yuanjun{use single-colored scatterplots} in the experiment of \yuanjun{region} density comparison in the formal study to eliminate the color effect.


\section{Formal Study}
\label{sec:experiments}

\subsection{Hypotheses}
\looseness=-1
\doc{This} formal study aims to evaluate the performance of seven selected sampling strategies \doc{based on} preserving three identified visual factors, including relative density, outlier, and overall shape.
According to the three visual factors, we \doc{formulated} four hypotheses to guide the experiment design.
Specifically, we \doc{formulated} two hypotheses on relative density in terms of \emph{region density} and \emph{class density}, respectively.
\emph{Region density} refers to the density of data points regardless of class information.
\emph{Class density} is the density of data points belonging to a certain class.

\noindent\textbf{H1:}
\textbf{All other sampling strategies perform better than random sampling in preserving relative region density.}\\
Maintaining relative density is a common goal for many sampling strategies~\cite{Chen14, Chen20, Hu20}. Compared to random sampling, these strategies are designed with delicate algorithms.
They often report positive results when \doc{compared} with random sampling in different scenarios~\cite{Chen20}.
Therefore, we assume all other sampling strategies should perform better than random sampling in preserving relative region density.

\noindent\textbf{H2: Multi-class adapted sampling strategies perform better than other sampling strategies in preserving relative class density.}\\
Many sampling strategies \doc{are customized for multi-class scatterplots.}
In these strategies, preserving the individual class properties, \eg, density, is an important goal.
Therefore, we assume that multi-class adapted sampling strategies, including multi-view Z-order sampling~\cite{Hu20}, recursive subdivision based sampling~\cite{Chen20}, and multi-class blue noise sampling~\cite{Chen14}, perform better than the other four sampling strategies in preserving relative class density.


\noindent\textbf{H3:}
\looseness=-1
\textbf{Outlier biased density based sampling is the best in preserving outliers.}\\
Many sampling strategies have shown \doc{an} ability \doc{to preserve} outliers in their reports. Among them, outlier biased density based sampling is designed especially for preserving outliers. It is the only strategy that integrates outlier measures into the sampling process. Therefore, we assume that outlier biased density based sampling is the best strategy \doc{for} preserving outliers.



\noindent\textbf{H4:}
\textbf{Blue noise sampling \yuanjun{and multi-class blue noise sampling perform} better than other strategies in preserving the overall shape.}\\
In our observation, uniform distribution facilitates the description of shapes by minimizing the effects of other visual factors, such as outliers and inhomogeneous density.
Based on \doc{this} observation, we assume that \yuanjun{sampling strategies that aim to generate uniform samples with blue noise property, \yuanjun{\ie}, blue noise sampling and multi-class blue noise sampling}, should perform the best in preserving the overall shape.

\subsection{Experiments}
Guided by the four hypotheses (\textbf{H1}--\textbf{H4}), we designed four experiments:
Experiment 1 (\textbf{E1}) was designed for the perception of relative region density preservation (\textbf{H1}), and Experiment 2 (\textbf{E2}) was designed for the perception of relative class density preservation (\textbf{H2});
Experiment 3 (\textbf{E3}) was designed for the perception of outlier maintenance (\textbf{H3});
Experiment 4 (\textbf{E4}) was for the perception of overall shape preservation (\textbf{H4}).
Note that \textbf{E1}--\textbf{E3} \doc{were} controlled experiments, and \textbf{E4} \doc{was} a subjective experiment.

\begin{figure}[!b]
\centering
\vspace{-5mm}
\includegraphics[width=\linewidth]{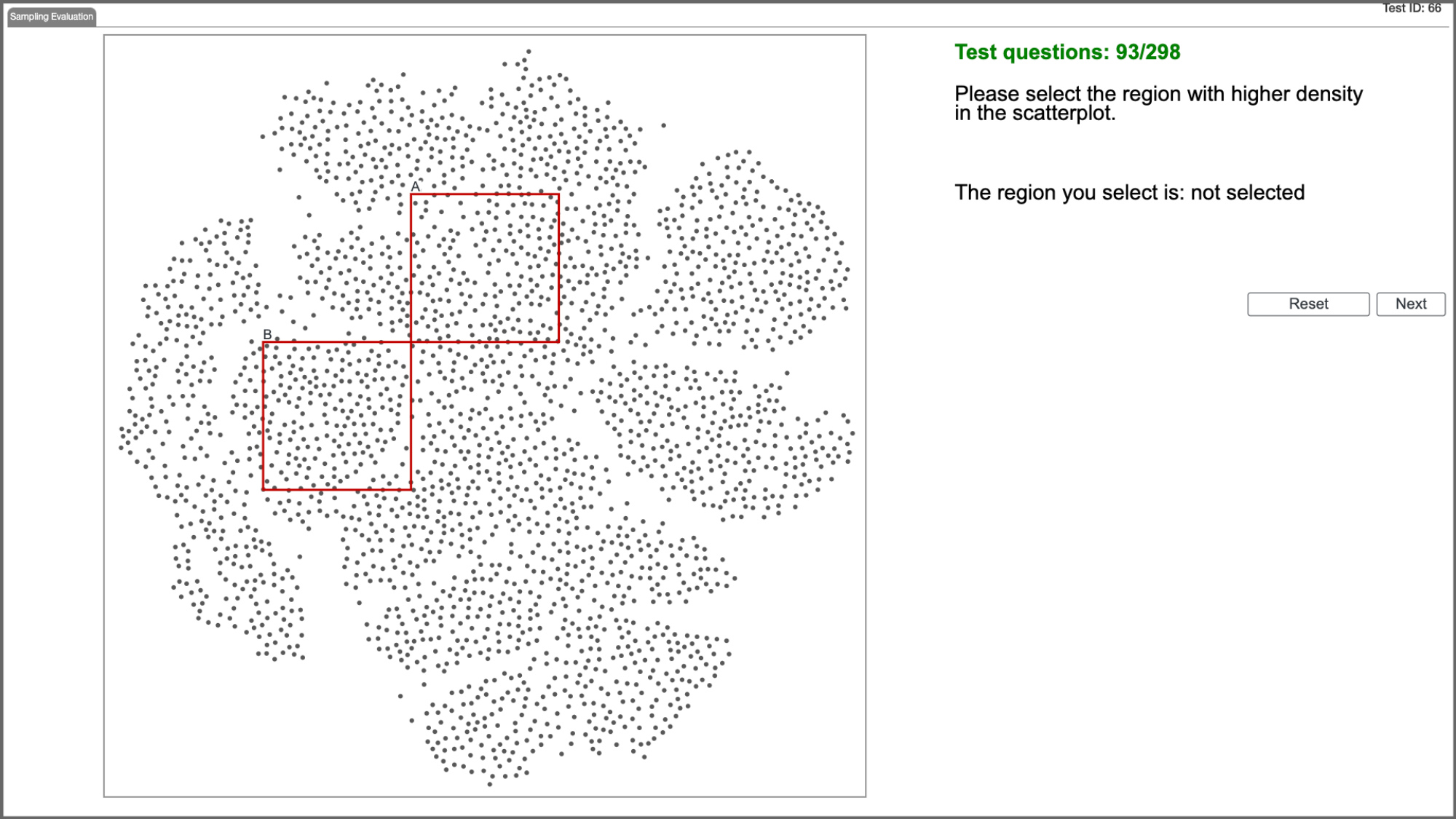}
\vspace{-5mm}
\caption{Example interface of \textbf{E1} in the formal study.}
\vspace{-5mm}
\label{fig:interface_E1}
\end{figure}

\noindent\textbf{E1: Perception of relative \yuanjun{region} density preservation}.
This experiment was used to evaluate the ability of different sampling strategies \doc{to preserve} relative region density in the aspect of visual perception.
Specifically, in \doc{this} experiment, we aimed to test if the region with higher \yuanjun{region} density can still be recognized as the higher one after sampling.
Thus, in each question, we randomly marked out two rectangle regions with the size of $\frac{w}{5}\times\frac{w}{5}$, where $w$ is the width of scatterplots.
Participants were asked to select the region with \doc{a} higher density \yuanjun{without considering class labels}.

Based on the result of the pre-study, color would interfere and slow down the judgments of the participants.
Thus, we rendered all data points in dark grey regardless of their labels.
We had eight datasets, and we generated two questions for each dataset.
For each question, we generated seven trials corresponding to seven sampling strategies, respectively.
In the seven trials of the same question, the locations of the rectangle regions were the same.
In total, we had
$$
\setlength{\abovedisplayskip}{3pt}
\setlength{\belowdisplayskip}{3pt}
    7\ (sampling\ strategies) \times 8\ (datasets) \times 2\ (questions) = 112
$$
trials for each participant.



\noindent\textbf{E2: Perception of relative \yuanjun{class} density preservation.}
This experiment was used to test whether the class \doc{with} a higher density can still be recognized as the higher one after sampling.
In contrast to \textbf{E1}, \textbf{E2} focuses on preserving relative density of specific classes \yuanjun{in the same region} instead of relative \yuanjun{region} density.
Thus, the scatterplots are rendered using color to encode class labels.
Specifically, in each question, we marked out a rectangle region with the size of $\frac{w}{5}\times\frac{w}{5}$ in a scatterplot.
We specified two classes in the question, and the participants were asked to choose the class with higher average density in the marked region.
\revision{In order to ensure fairness, the color assigned to each class in a specific dataset was identical across all sampling strategies.}
Similar to \textbf{E1}, we generated two questions for each dataset and seven trials for each question. In total, we had
$$
\setlength{\abovedisplayskip}{3pt}
\setlength{\belowdisplayskip}{3pt}
    7\ (sampling\ strategies) \times 8\ (datasets) \times 2\ (questions) = 112
$$
trials for each participant.\looseness=-1



\noindent\textbf{E3: Perception of outlier maintenance.}
This experiment was used to evaluate the ability of sampling strategies in preserving outliers in the aspect of perception.
We tested: (1) whether an outlier in the original dataset can still be preserved and perceived as an outlier after sampling; and (2) in case a point is perceived as an outlier after sampling, whether it is indeed an outlier in the original dataset.



There are diverse definitions of outliers, however, here we focused on the outliers in two \doc{situations}: first, when considering class labels, the point that is of \doc{a} different class \doc{from} its neighboring points; second, when not considering class labels, the points which are located \yuanjun{at abnormal distances} from its class.
We followed the definition in the class purity algorithm~\cite{Motta2015} in the first scenario and followed the local outlier factor algorithm ~\cite{pedregosa2011scikit} in the second scenario.

\looseness=-1
\yuanjun{In each question, we marked out a region in a scatterplot.}
Participants were asked to mark the outliers \doc{assuming} they \doc{believed} the outliers existed in the marked region.
Note that, the outliers were referring to all points in the entire scatterplot, instead of the marked region.
We \doc{considered} three ways for participants to identify outliers:
first, marking out all the outliers in the entire scatterplot;
second, marking out a specified number of outliers (\eg, 10) in the entire scatterplot;
and third, marking out all outliers in a given rectangle region.
Considering the huge \doc{number} of outliers in the entire dataset, it is not feasible to mark all \doc{of them} in the entire range in the limited experiment time.
\doc{In addition}, the accuracy would be very low since \doc{many} outliers would be missed.
If we limit the number of target outliers as noted in the second option, due to the large number of outliers, participants may easily mark \doc{the} requested number of outliers.
The accuracy would be high for all \doc{the} strategies, and it would be hard to distinguish the performances of different sampling strategies,
\doc{so} we chose the third option and asked the participants to mark all the outliers in a fixed range.
The only disadvantage \doc{with} this option was that participants might mis-select outliers referring to the local distribution.
To avoid this, we reminded the participants to refer to global distribution, and we also corrected the observed errors in the training session.
In total, we had
$$
\setlength{\abovedisplayskip}{3pt}
\setlength{\belowdisplayskip}{3pt}
    7\ (sampling\ strategies) \times 8\ (datasets) = 56
$$
trials for each participant in this experiment.

\looseness=-1
\noindent\textbf{E4: Perception of overall shape preservation.}
This experiment was used to compare the abilities of sampling strategies \doc{to preserve} the overall shape of scatterplots in terms of visual perception.
In contrast to \textbf{E1}--\textbf{E3}, \textbf{E4} was a subjective experiment.
In each trial, we sampled a dataset using seven strategies and displayed these seven sampling results together with the original scatterplot (see Fig.~\ref{fig:interface_1}).
Participants were asked to rank the seven sampling results based on the shape similarities between the sampling results and the original scatterplot.
Participants were reminded that class labels should be taken into account in comparing the shape similarities.
Parallel rankings were allowed when participants could not distinguish the difference among the sampling results.
\revision{Each participant had one trial for each dataset.
Thus, in total, we had eight trials for each participant in this experiment.}


\subsection{Participants, Apparatus and Testing Data}

\noindent\textbf{Participants}.
We recruited 100 participants (78 males, 22 females, aged 18--50 years, average: 24) for the formal study.
16 of them are researchers in visualization and computer graphics.
The others are undergraduates or graduated students majoring in computer science.
34 participants reported previous experience with sampling.
None of them reported color blindness or color weakness.
Each participant was rewarded \$20 per hour for completing the experiments.

\noindent\textbf{Apparatus}.
\revision{The experiments were conducted online through a web prototype (see Figure \ref{fig:interface_E1}).}
Participants were required to visit it remotely on the Chrome browser and finish the experiment on a screen with a resolution of $1,920 \times 1,080$.
They were asked to share their screen with the instructor during the experiments to enable remote monitoring.

\noindent\textbf{Testing data}.
We generated scatterplots based on the \doc{eight selected} datasets for the experiments \revision{in advance}.
For each dataset, we created one scatterplot of the original dataset and seven scatterplots of sampling results by the seven sampling strategies, respectively.
The sampling rates of each dataset were determined \doc{based on} the results of the pre-study.
Since multi-view Z-order sampling and recursive subdivision based sampling cannot set the exact sampling rate, we controlled the error \doc{at} $1\%$.
\yuanjun{The points were rendered with a radius of 3 pixels without transparency.
The size of the scatterplots was $1,000 \times 1,000$ pixels in \textbf{E1}--\textbf{E3}, and $300 \times 300$ pixels in \textbf{E4}.}
In order to avoid imbalanced occlusion between classes, the points in the scatterplots were rendered in random order.
Except for \textbf{E1}, we selected Boynton's color palette~\cite{Boynton1989} to encode classes in the scatterplots.
For the training session, we generated synthetic datasets following the Gaussian mixed distribution.
In the real testing session, the order of all \doc{the} questions was counterbalanced by following a Latin square to avoid the learning effect.

\subsection{Procedure}

Each experiment included a training session and a real test session.
At the beginning of the training session, the instructor explained the experiments as well as the related concepts (\eg, outliers).
After the explanation, several practice trials (three for \textbf{E1}--\textbf{E3}, and one for \yuanjun{\textbf{E4}}) were presented to help participants get familiar with the experiments.
For controlled experiments, \textbf{E1}--\textbf{E3}, the correct answers were shown to the participants after the answers were submitted.
The participants were encouraged to ask questions in the training sessions to facilitate their understanding of the experiments.
After they reported that they have fully understood the tasks, we started the real test session.
Participants were allowed to have a break of five minutes before each experiment.
After the participants completed all the experiments, they were asked to
answer a questionnaire for their backgrounds and subjective feedback on the experiments.
The entire process lasted approximately one hour and 20 minutes for each participant.
During the online experiments, network error occurred in nine trials, resulting in unusual long response time (more than 30 seconds). Considering that repeating these nine trials will also introduce bias into the result, we simply discarded these nine trials to preserve the validity of result.

\looseness=-1
The questionnaire included three parts.
First, we asked participants \doc{about their} backgrounds and basic information, familiarity with visualization, and experience with sampling strategies.
Second, participants were asked to rate the importance of
preserving relative density, outliers, and overall shape for a sampling strategy using a five-point Likert scale.
They were also encouraged to add extra abilities that a sampling method should provide.
\doc{Finally}, we asked \doc{about} their focus in each experiment in order to learn the important visual factors for human perception.















\begin{figure}[!b]
\centering
\vspace{-5mm}
\includegraphics[width=\linewidth]{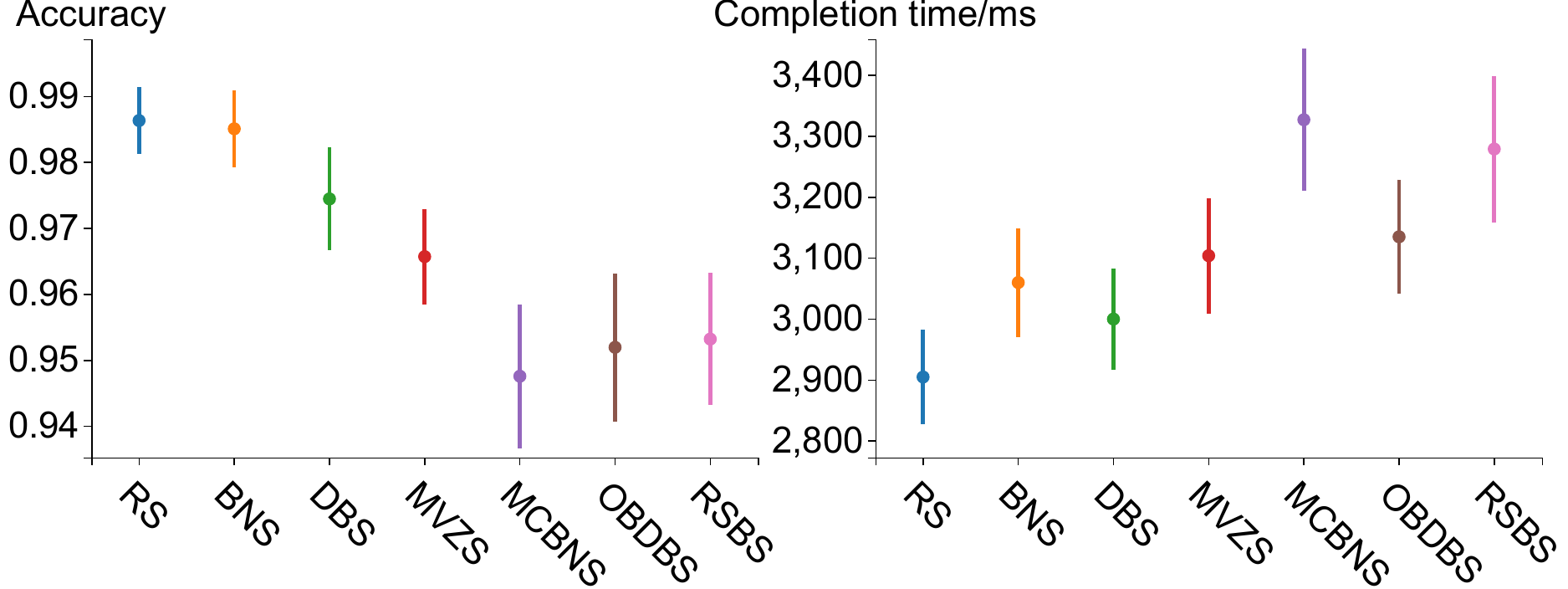}
\vspace{-6mm}
\caption{Average accuracy and completion time of \textbf{E1} (Perception of relative region density preservation).
}
\label{fig:formal_task1}
\end{figure}

\section{Experimental Results}
\label{sec:results}


\subsection{Analysis Approach}
We recorded the objective measurements from \textbf{E1}--\textbf{E3} and the subjective measurement from \textbf{E4}.
For \textbf{E1} and \textbf{E2}, we recorded the correctness and completion time of each trial.
For \textbf{E3}, we calculated the precision and recall of each trial.
In each trial, we denoted the set of outliers marked out by a participant as $M$ and the ground truth as $N$.
The precision is the ratio of $|N \cap M|$ \jiazhi{to} $|M|$, and the recall is the ratio of $|N \cap M|$ \jiazhi{to} $|N|$, where $| \cdot |$ denotes to the cardinality of a finite set.
Note that the recall refers to the ratio of outliers that are preserved by sampling and then perceived by participants.
To avoid small values, we normalized the recall by the maximal outlier preserving ratio among seven sampling strategies on each dataset.
Without \yuanjun{loss} of clarity, we use the term \emph{recall} to refer to normalized recall in the rest of this paper.
For \textbf{E4}, we recorded the ranking of each sampling strategy in each trial and \doc{transformed} the rankings into scores.
Specifically, the 1st--7th sampling strategies get 7--1 point, respectively.


\looseness=-1
Following the common \doc{methods for} evaluating user performance~\cite{lam2012empirical}, we reported the mean values and confidence intervals of the objective measurements and performed significance analyses to test our hypotheses.
For the results of \textbf{E1}--\textbf{E3}, we performed \doc{the} Shapiro-Wilk test and found that they do not follow the normal distribution.
Therefore, we employed a non-parametric method to examine whether significant differences exist among the sampling strategies.
Specifically, we chose the Friedman test with the standard significance level \doc{of} $\alpha = 0.05$ in our analysis.
If there \doc{were} significant differences, we conducted \doc{the} Conover test as the post-hoc test to examine the pairwise significance.
In \textbf{E4}, we also reported the mean value and the confidence interval of the rating score for each sampling strategy.

\begin{figure}[!b]
\centering
\vspace{-5mm}
\includegraphics[width=\linewidth]{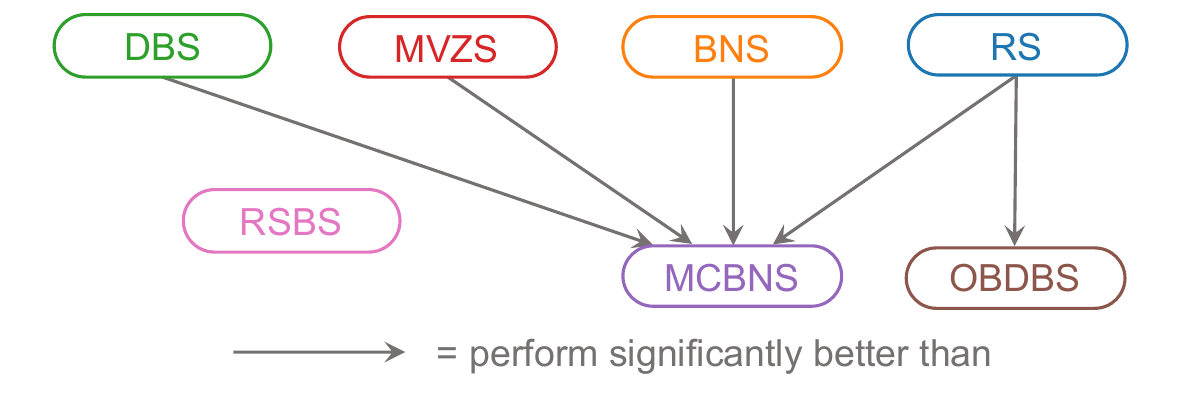}
\vspace{-8mm}
\caption{Graphical depiction of the pairwise significance relationships of the accuracy differences of the sampling strategies in \textbf{E1}. A directed edge indicates that the origin sampling strategy performs significantly better than the destination one. Same as below.}
\label{fig:DAG_task1_acc}
\end{figure}

\begin{figure}[!b]
\centering
\vspace{-5mm}
\includegraphics[width=\linewidth]{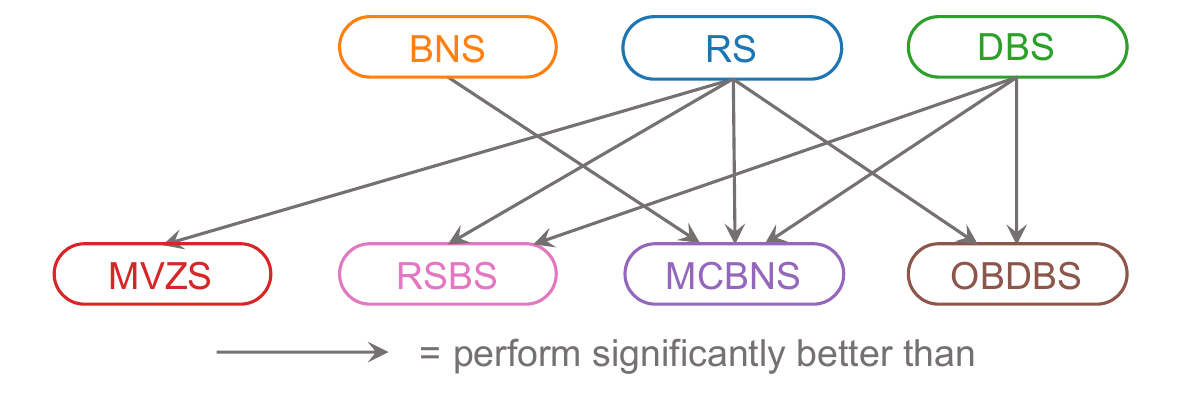}
\vspace{-8mm}
\caption{Graphical depiction of the pairwise significance relationships of the completion time differences of the sampling strategies in \textbf{E1}.}
\label{fig:DAG_task1_time}
\end{figure}

\subsection{\yuanjun{Results Analysis}}
\noindent\textbf{H1:}
\looseness=-1
We assume that all other sampling strategies perform better than random sampling in preserving relative region density.

\revision{
\textbf{H1} is rejected as random sampling performs the best in preserving relative region density.
}

\begin{figure}[t]
\centering
\includegraphics[width=\linewidth]{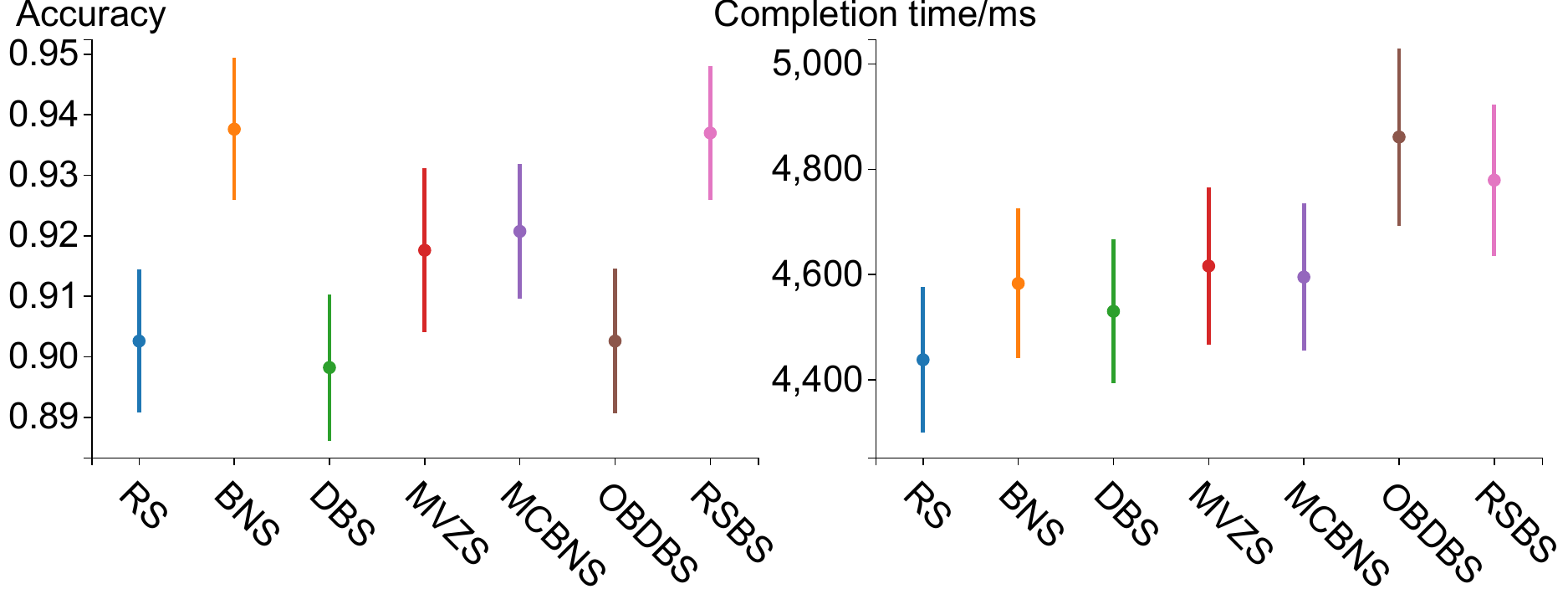}
\vspace{-6mm}
\caption{Average accuracy and completion time of \textbf{E2} (Perception of relative class density preservation).
}
\vspace{-3mm}
\label{fig:formal_task2}
\end{figure}

\begin{figure}[t]
\centering
\includegraphics[width=\linewidth]{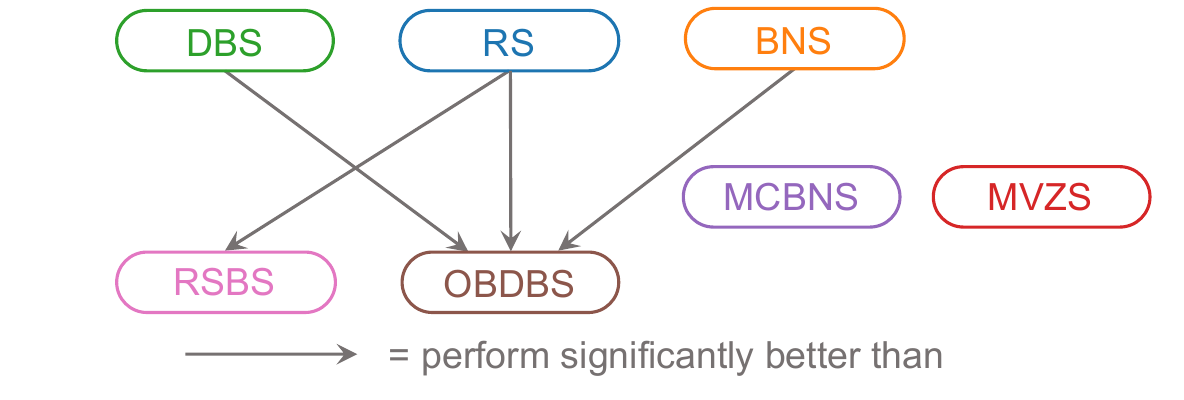}
\vspace{-8mm}
\caption{Graphical depiction of the pairwise significance relationships of the completion time differences of the sampling strategies in \textbf{E2}.}
\label{fig:DAG_task2}
\vspace{-7mm}
\end{figure}

\begin{figure}[t]
\centering
\includegraphics[width=\linewidth]{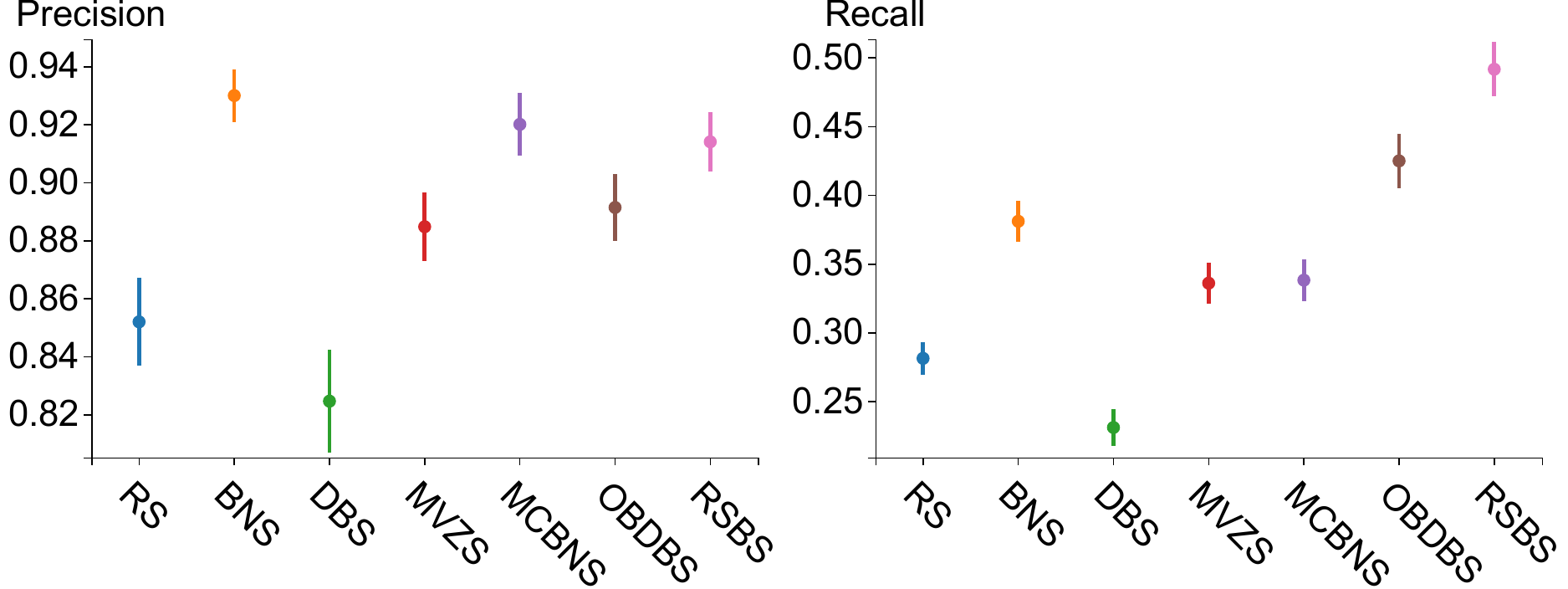}
\vspace{-6mm}
\caption{Precision and recall of \textbf{E3} (Perception of outlier maintenance).
}
\vspace{-3mm}
\label{fig:formal_task3}
\end{figure}

\begin{figure}[t]
\centering
\includegraphics[width=\linewidth]{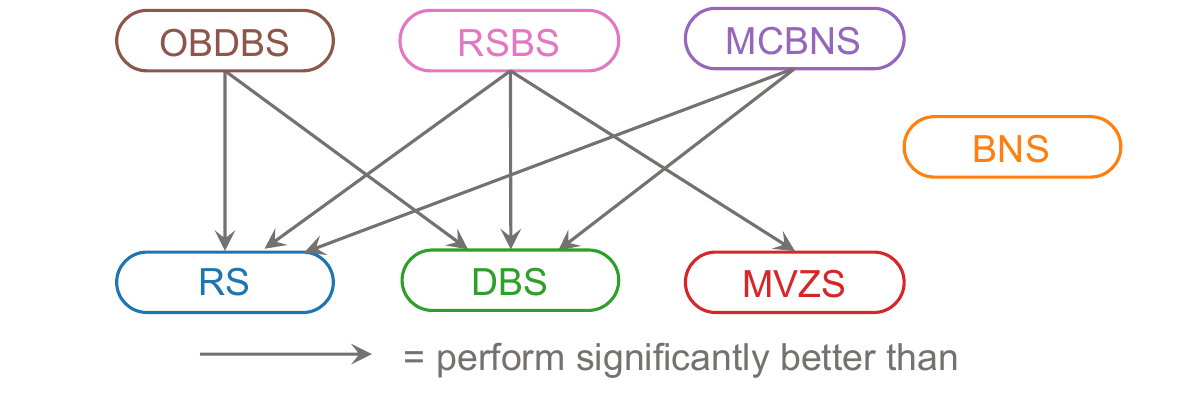}
\vspace{-8mm}
\caption{Graphical depiction of the pairwise significance relationships of the recall differences of the sampling strategies in \textbf{E3}.}
\vspace{-7mm}
\label{fig:DAG_task3}
\end{figure}

Fig.~\ref{fig:formal_task1} shows the results of \textbf{E1}.
\revision{Among all sampling strategies, the accuracy ranges from $94.8\%$ to $98.7\%$, while the average completion time ranges from 2,855ms to 3,283ms.}
Random sampling has the highest accuracy ($98.63\%$) \yuanjun{and the shortest completion time ($2904ms$)} in \textbf{E1}.
The Friedman tests show that statistical significance \yuanjun{among} different sampling strategies exist in terms of accuracy ($\chi^2(6)=13.56, p = 0.0349$) and average completion time ($\chi^2(6)=20.28, p = 0.0025$).
Fig.~\ref{fig:DAG_task1_acc} \yuanjun{depicts} the pairwise significance relationships between each pair of sampling strategies in terms of accuracy.
Random sampling performs significantly better than multi-class blue noise sampling ($p = 0.0051$) and outlier biased density based sampling ($p = 0.0232$) in terms of accuracy.
Fig.~\ref{fig:DAG_task1_time} \yuanjun{depicts} the pairwise significance relationships in terms of average completion time.
Random sampling performs significantly better than multi-view Z-order sampling ($p = 0.0328$), multi-class blue noise sampling ($p = 0.0011$), outlier biased density based sampling ($p = 0.0221$), and recursive subdivision based sampling ($p = 0.0048$) in terms of average completion time.
No sampling strategy performs significantly better than random sampling, either in terms of accuracy or average completion time.

\noindent\textbf{H2:}
We assume that multi-class adapted sampling strategies, including multi-class blue noise sampling, multi-view Z-order sampling, and recursive subdivision based sampling, perform better than random sampling, blue noise sampling, \yuanjun{density biased sampling} and outlier biased density based sampling in preserving relative class density.

\revision{
\textbf{H2} is partially confirmed as multi-class sampling strategies achieve higher accuracy except for blue noise sampling, while random sampling performs the best in terms of completion time.
}

The results of \textbf{E2} are displayed in Fig.~\ref{fig:formal_task2}.
Blue noise sampling has the highest accuracy at $93.75\%$.
The accuracies of recursive subdivision based sampling ($93.69\%$), multi-class blue noise sampling ($92.06\%$), and multi-view Z-order sampling ($91.75\%$) are higher than the remaining three strategies.
\revision{With an accuracy range of nearly $4\%$ ($89.81\%$--$93.75\%$),}
\doc{however, no significant difference in accuracy is reported using the Friedman test} ($\chi^2(6)=4.019, p = 0.6741$).

\looseness=-1
In \doc{terms} of average completion time, random sampling performs the best,
\revision{obtaining a result of 4,437ms, while the worst one, outlier biased density based sampling, is more than 400ms slower.}
The three multi-class adapted sampling strategies \doc{offer} no \doc{clear} advantage \doc{compared to} other strategies.
The Friedman test shows that \doc{a} significant difference exists in completion time ($\chi^2(6)=12.78, p = 0.0467$).
Fig.~\ref{fig:DAG_task2} shows the pairwise significance relationships in terms of average completion time.
Random sampling performs significantly better than recursive subdivision based sampling ($p = 0.0221$).
Besides testing the hypothesis, we also find that outlier biased \yuanjun{density based} sampling performs significantly worse than density biased sampling \yuanjun{($p = 0.0270$)}, random sampling \yuanjun{($p = 0.0095$)}, and blue noise sampling \yuanjun{($p = 0.0270$)}.

\begin{figure*}[ht]
\centering
\includegraphics[width=\linewidth]{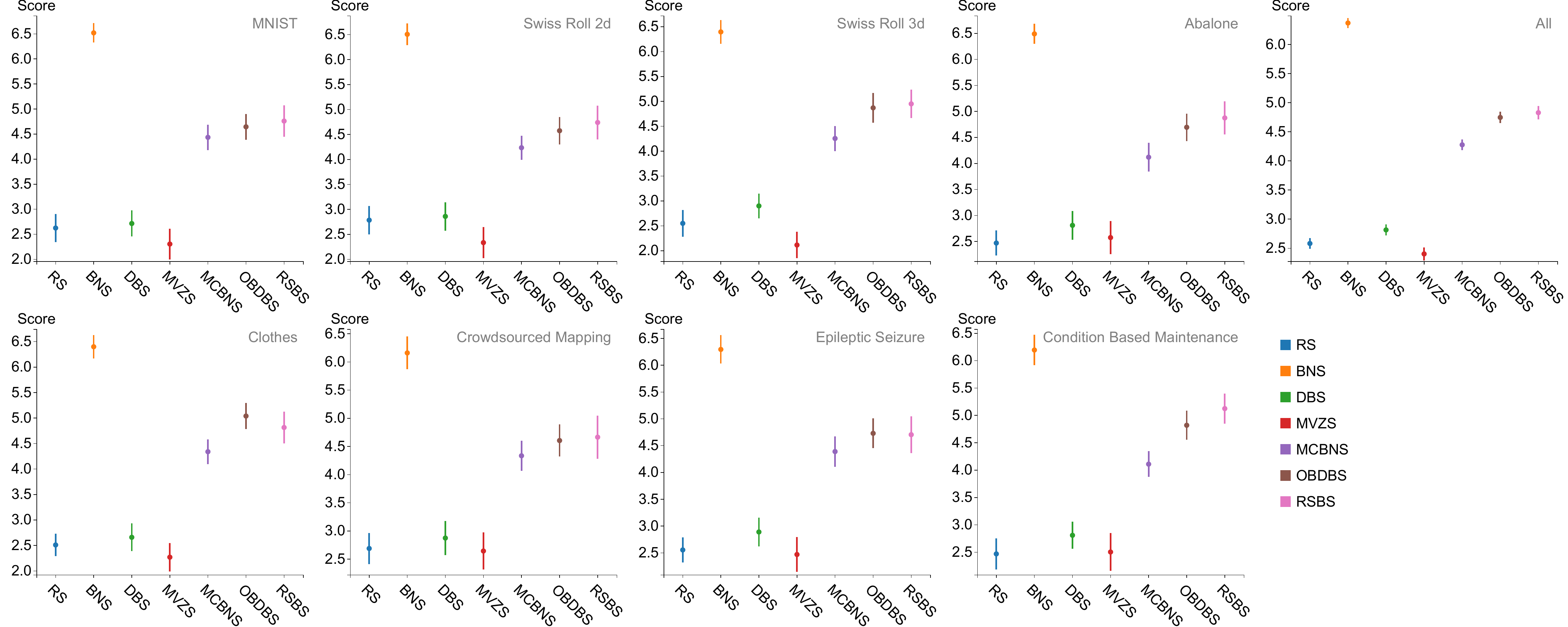}
\vspace{-5mm}
\caption{Average ranking scores of \textbf{E4} (Perception of overall shape preservation).
}
\vspace{-5mm}
\label{fig:formal_task4}
\end{figure*}

\noindent\textbf{H3:}
We assume that outlier biased density based sampling is the best in preserving outliers.

\revision{
\textbf{H3} is partially confirmed as recursive subdivision based sampling and outlier biased density based sampling achieve higher recall than other strategies, while blue noise sampling has the highest precision.
}

As shown in Fig.~\ref{fig:formal_task3}, outlier biased density based sampling is \doc{ranked fourth} in terms of precision.
Blue noise sampling, multi-class blue noise sampling, and recursive subdivision based sampling have higher precision than outlier biased density based sampling.
\revision{The precision data span more than $10\%$ ($82.45\%$--$92.98\%$) among all strategies.}
However, the Friedman test \doc{finds} no significant differences in \doc{the precision} of the sampling \yuanjun{strategies} ($\chi^2(6) = 10.53, p = 0.1040$).
In \doc{terms} of recall, outlier biased density based sampling is ranked second, while recursive subdivision based sampling has the highest recall.
\revision{The range of recall is about $26\%$, from $23.08\%$ to $49.12\%$, and}
the Friedman test shows that significant differences exist ($\chi^2(6) = 18.78,  p = 0.0045$).
The post-hoc tests show that outlier biased density based sampling performs significantly better than random sampling and density based sampling (Fig.~\ref{fig:DAG_task3}).
Fig.~\ref{fig:DAG_task3} shows the discovered pairwise significance relationships.
\doc{In addition to} the hypothesis test, we also \doc{find} some \doc{other} interesting \doc{results}.
First, blue noise sampling has the highest precision and the third-highest recall.
Although it has no significance relationship with other strategies in terms of precision and recall, it is worth \doc{recommending it for} outlier preservation \yuanjun{along with outlier biased density based sampling and recursive subdivision based sampling}.
In contrast, random sampling and density biased sampling have relatively lower precision and recall than \doc{the} other strategies.
Significance difference\lingyun{s} \doc{are found to} exist between them and outlier biased density based sampling, multi-class blue noise sampling, and recursive subdivision based sampling.

\noindent\textbf{H4:}
We assume that blue noise sampling and multi-class blue noise sampling perform better than other strategies in preserving the overall shape.

\revision{
\textbf{H4} is also partially confirmed as blue noise sampling gets the highest score, but multi-class blue noise sampling only ranks 4th.
}

Fig.~\ref{fig:formal_task4} shows the average ranking scores of sampling strategies in eight datasets and their average.
Blue noise sampling has the highest ranking score in all eight datasets \yuanjun{with} an average score of $6.37$, \yuanjun{while the performance of multi-class blue noise sampling is \doc{near the middle}}.
\yuanjun{Moreover,} recursive subdivision based sampling, outlier biased density based sampling, and multi-class blue noise sampling has similar \doc{scores} and \doc{rank} 2nd, 3rd, and 4th \yuanjun{with \doc{averages} of $4.82$, $4.74$, and $4.27$}, respectively.
The ranking is stable across \doc{all} eight datasets.

\noindent
\revision{
\textbf{Takeaways}.
Since blue noise sampling has competitive results in all experiments, it is suggested to be more generally used in data exploration.
Random sampling performs comparatively well in both E1 and E2, indicating that it is still a competitive choice when users seek to preserve the relative density in the sampled scatterplot given its simplicity.
In addition, as outlier biased density based sampling and recursive subdivision based sampling show their capabilities in outlier maintenance and shape preservation, users may pay more attention to them when encountering such practical needs.
}

\section{Discussion}
\label{sec:discussion}

\subsection{Important Visual Factors in Sampling}
\begin{figure}[t]
\centering
\includegraphics[width=\linewidth]{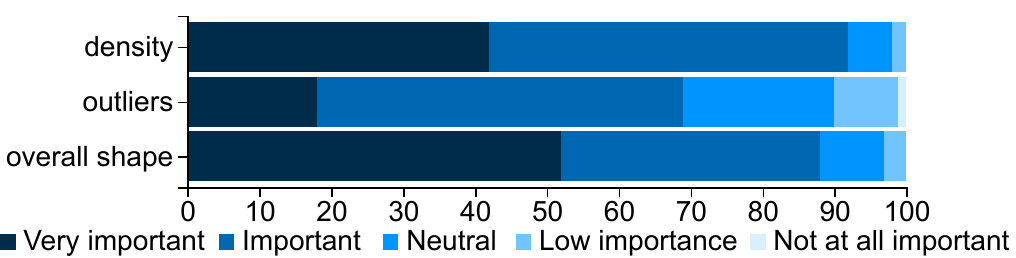}
\vspace{-6mm}
\caption{The importance rating of the three evaluated visual factors.}
\vspace{-6mm}
\label{fig:visual_factor}
\end{figure}

\looseness=-1
The subjective questionnaire on the \doc{important} visual factors provided insights into sampling strategy selection in different scenarios.
The results are presented in Fig.~\ref{fig:visual_factor}.
The average ratings of relative density, outliers, and overall shape, are 4.32, 3.76, and 4.37, respectively.
The high ratings confirm that the evaluated visual factors are \doc{common concerns} in scatterplot sampling.
We also \doc{got} interesting findings when considering the difference between participants from different fields.
\doc{The} participants in visualization and computer graphics rated 4.00 for outliers and 4.23 for overall shape.
Compared to the averages, they \doc{were} more interested in outlier maintenance than \doc{the} participants in other fields.
In contrast, \doc{the} participants in computer vision and deep learning rated 3.39 for outliers, \doc{but} 4.69 for overall shape.
The participants commented that they \doc{were} particularly interested in classification and regression tasks, and the overall shape is helpful \yuanjun{in understanding} the pattern of classification and correlation.
The participants in computer graphics paid more attention to relative density preservation with a rating of 4.63.
\doc{This} is because density is important in geometry modeling tasks of computer graphics.
Therefore, we can recommend sampling strategies according to the specific task.
For instance, although people in computer vision and deep learning are not familiar with blue noise sampling, they \lingyun{are} suggested to try it because blue noise sampling performs the best in overall shape preservation.
In addition, extra visual factors \doc{were} proposed in the subjective questionnaire.
\lingyun{For instance, }\jiazhi{the position relationship among classes \yuanjun{(\eg, whether two classes are separated or mixed together)} \doc{was} proposed by 15 participants. }
Three participants commented that the trends \doc{regarding} points should also be preserved.
These ideas might shed light on new design requirements for sampling.

\subsection{Influencing Factors of Perception and Design Considerations for Sampling}
\looseness=-1
The subjective questionnaire also contained factors that affect perceptions during the experiments.
These results provided insights \doc{into} \doc{possible revisits of} our experiment design and shed light on sampling strategy design for different tasks.
In \textbf{E1}, the covering area (by 60\% of the participants) and the distance among points (by 50\% of the participants) \doc{were} reported to affect region density judgments.
\yuanjun{The occupancy model~\cite{Allk1991} shows that how these factors affect human perception of region density on scatterplots, and there are some previous works~\cite{Vos1988, Bertini2006} \doc{that measured the} perceptual density difference of some factors through user studies.}
\yuanjun{Bertini~\etal~\cite{Bertini2006} leverage an ad-hoc perception study result and propose a sampling framework to strengthen the perception of relative density differences.}
\yuanjun{Further exploration \doc{of} integrating perceptual effects with sampling strategy design will be an interesting direction.}
In \textbf{E2}, the covering area, the distance among points, and the colors of two classes (25\%) \doc{were} reported to be effective visual factors \doc{when making} class density judgments.
\yuanjun{Specifically, }\yuanjun{56\%} of \doc{the respondents} commented that bright color leads to over-estimation of the density.
In our experiment, we randomly set colors from Boynton's color palette~\cite{Boynton1989} \doc{that} are considered to be almost never confused.
The color issue may be further alleviated by optimizing the color assignments to classes after sampling like Wang~\etal~\cite{wang2019optimizing}.
In \textbf{E4}, \yuanjun{66\%} of participants commented that the outline of the shape \doc{was} the most important visual factor when comparing overall shapes.
Inspired by this report, a sampling strategy \doc{aimed} at overall shape preservation should pay more attention to the boundary of clusters.

\subsection{Limitations and Future Work}
\looseness=-1
As mentioned above, there are many factors affecting the perception of the evaluated visual factors.
On the one hand, these factors, for example, the color in \textbf{E2}, may introduce perceptual bias in our experiments.
Considering that we have a \doc{large} number of trials, \yuanjun{such} bias can be reduced by random settings for the trails.
On the other hand, understanding the relationships between them and \doc{the} visual factors \doc{we are concerned with} would be inspiring for \doc{future} sampling strategy design.
However, a controlled experiment containing all \doc{of} these variables would \doc{make it hard to \lingyun{conduct}} a practical evaluation.
It would be interesting to perform further evaluation of the relationships among them.


\looseness=-1
In addition, our evaluation only considered sampling on 2D data.
However, scatterplots are usually employed to visualize high-dimensional data \doc{in conjunction with} dimensionality reduction approaches.
Sampling is performed in the high-dimensional space rather than 2D space for efficiency.
A promising future direction is to explore the perception effects of sampling strategies when they are performed in the high-dimensional space.



\section{Conclusion}
\label{sec:conclusion}

\looseness=-1
In this paper, we present an empirical evaluation of sampling strategies for scatterplots from the perspective of perception.
We identify seven representative sampling strategies and three critical visual factors for scatterplots following a comprehensive survey \doc{of} the existing literature.
Based on the results, we formulate four hypotheses and design four experiments to evaluate the ability of the selected sampling strategies to preserve the identified visual factors.
We first conduct a pre-study to determine the proper sampling number of each dataset and confirm the negative effect on region density identification caused by color.
The results of the formal study show that (1) random sampling is the best in region density preservation in terms of time and accuracy; (2) blue noise sampling and multi-class sampling strategies \doc{are accurate at} class density preservation, while random sampling \doc{is highly efficient at} this task; (3) recursive subdivision based sampling, outlier biased density based sampling, and blue noise sampling are favored in outlier maintenance; \doc{and} (4) blue noise sampling is the best in overall shape preservation.
These results offer practical guidance for the selection of sampling strategies in different application scenarios.

\acknowledgments{
This research is supported by the National Key R\&D Program of China (No.s 2018YFB1004300, 2019YFB1405703), the National Natural Science Foundation of China (No.s 61761136020, 61672307, 61672308, 61872389, 61936002), XJTLU Research Development Funding RDF-19-02-11, and TC190A4DA/3.}


\small
\bibliographystyle{abbrv}
\bibliography{reference}


\end{document}